\documentclass[letterpaper,english,aps, reprint]{revtex4-1}
\usepackage[T1]{fontenc}
\usepackage[latin9]{inputenc}
\usepackage[active]{srcltx}
\usepackage{amsmath}
\usepackage{amssymb}
\usepackage{graphicx}
\usepackage[hyperindex,breaklinks]{hyperref}

\makeatletter

\usepackage{epstopdf}

\makeatother

\usepackage{babel}
\begin{document}
\preprint{APS/123-QED}
\title{A Kinetic Model for Electron-Ion Transport in Warm Dense Matter}
\author{Shane Rightley}
\email{shane-rightley@uiowa.edu}

\affiliation{Department of Physics and Astronomy~\\
 University of Iowa~\\
Iowa City, IA 52242 USA}
\author{Scott D. Baalrud}
\email{baalrud@umich.edu}

\affiliation{Department of Nuclear Engineering and Radiological Sciences~\\
 University of Michigan~\\
 Ann Arbor, MI 48109 USA}
\date{\today}
\begin{abstract}
We present a model for electron-ion transport in Warm Dense Matter
that incorporates Coulomb coupling effects into the quantum Boltzmann
equation of Uehling and Uhlenbeck through the use of a statistical
potential of mean force. Although this model has been derived rigorously
in the classical limit {[}S.D. Baalrud and J. Daligault, Physics of
Plasmas \textbf{26,} 8, 082106 (2019){]}, its quantum generalization
is complicated by the uncertainty principle. Here we apply an existing
model for the potential of mean force based on the quantum Ornstein-Zernike
equation coupled with an average-atom model {[}C. E. Starrett, High
Energy Density Phys. \textbf{25}, 8 (2017){]}. This potential contains
correlations due to both Coulomb coupling and exchange, and the collision
kernel of the kinetic theory enforces Pauli blocking while allowing
for electron diffraction and large-angle collisions. By solving the
Uehling-Uhlenbeck equation for electron-ion relaxation rates, we predict
the momentum and temperature relaxation time and electrical conductivity
of solid density aluminum plasma based on electron-ion collisions.
We present results for density and temperature conditions that span
the transition from classical weakly-coupled plasma to degenerate
moderately-coupled plasma. Our findings agree well with recent quantum
molecular dynamics simulations.
\end{abstract}
\maketitle

\section{\label{sec:introduction}Introduction}

The microscopic physics of WDM is subject to a multitude of physical
effects, including electron degeneracy, partial ionization, large-angle
scattering, diffraction, and moderate Coulomb coupling leading to
correlations. Such conditions are present in experiments involving
extreme compression of materials \citep{Glenzer2016,Riley2018,Mancic2010},
in astrophysics \citep{Redmer2008,Koenig2005}, and along the compression
path in inertial confinement fusion (ICF) experiments \citep{Hu2015}.
As a result of the demanding conditions for theoretical modeling,
the description of WDM has been highly reliant on computational techniques.
However, ab initio computation proves too expensive for many problems,
whereas faster methods often involve uncontrolled approximations or
have uncertain applicability. In order to support computational efforts,
explore larger regions of parameter space, and expediently provide
data tables for hydrodynamic simulations, reliable and fast tools
for the computation of transport coefficients in WDM remain desirable.

\begin{figure}[!tph]
\includegraphics[width=8.55cm]{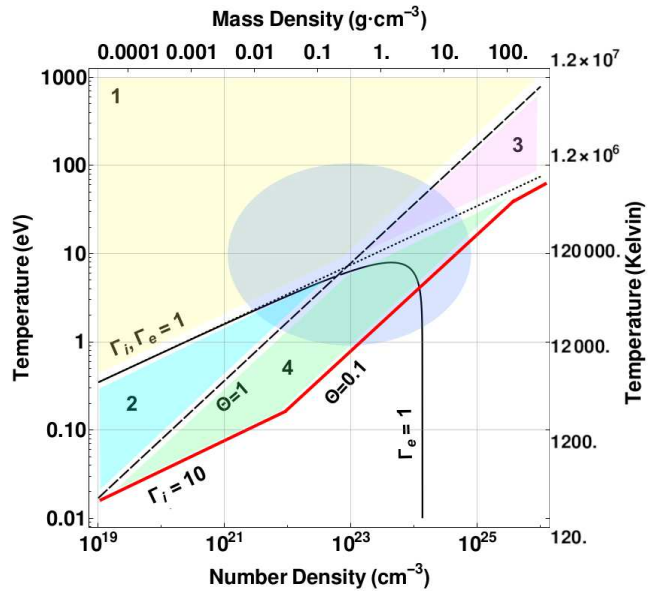}\caption{\label{fig:parregimes}Parameter regimes of fully ionized hydrogen
plasma. The solid black line is the boundary between weak and strong
electron coupling $\Gamma_{e}=1$ and turns over due to the electron
degeneracy; the dotted lined is the separation between weak and strong
ion coupling $\Gamma_{i}=1$; and the dashed line is the separation
between classical and degenerate electrons $\Theta=1$. The darker
blue oval denotes the sector of WDM. Region 1 (yellow) is classical
weakly coupled plasma; region 2 (light blue) is characterized by classical
strong coupling; region 3 (pink) by quantum weak coupling, and region
4 (green) by both quantum electrons and strongly-coupled ions. We
expect the theory presented here to apply to each region 1-4. The
red line demarcates the region of validity of plasma-type transport
theories; beyond this is the regime of condensed matter.}
\end{figure}

In this work, we introduce a model for electron-ion transport based
on the quantum Boltzmann equation of Uehling-Uhlenbeck \citep{Uehling1933},
but with a modification motivated by the classical mean force kinetic
theory \citep{Baalrud2019} in which aspects of many-body interactions
are modeled by treating binary collisions as occurring via the potential
of mean force. The model accounts for at least some degree of partial
ionization, electron degeneracy, moderate Coulomb coupling, diffraction,
and large-angle collisions. The approximate regimes in which these
different physical processes are important can be roughly understood
in terms of the degeneracy parameter $\Theta\equiv T_{e}/T_{F}$ and
Coulomb coupling parameter $\Gamma=\left\langle U\right\rangle /\left\langle K\right\rangle $
with the statistical averages taken using a Maxwell-Boltzmann distribution
for ions and a Fermi-Dirac distribution for electrons. $T_{e}$ is
the electron temperature, $T_{F}\equiv E_{F}/k_{B}$ the Fermi temperature,
$\left\langle U\right\rangle $ the average interaction energy and
$\left\langle K\right\rangle $ the average kinetic energy of a particle.
The average speed of electrons shifts from the thermal speed to the
Fermi speed as degeneracy increases, a phenomenon that causes electrons
to become increasingly weakly coupled at high density. The Coulomb
couplings $\Gamma_{ii}$ and $\Gamma_{ie}$ for ion-ion and electron-ion
interactions, respectively, can be expressed
\begin{equation}
\Gamma_{ii}=\frac{Z^{2}e^{2}/a}{k_{B}T},
\end{equation}
and
\begin{equation}
\Gamma_{ie}=\frac{Ze^{2}/a}{k_{B}T}\frac{{\rm Li}_{3/2}\left[-\xi\right]}{{\rm Li}_{5/2}\left[-\xi\right]},
\end{equation}
where $a=(3/4\pi n)^{1/3}$ is the Wigner-Seitz radius, ${\rm {\rm Li}}$
is the polylogarithm function (closely related to the Fermi integral)
and $\xi\equiv\exp(\mu/k_{{\rm B}}T)$ where $\mu$ is the electron
chemical potential related to $\Theta$ through the normalization
of the Fermi-Dirac distribution \citep{Melrose2010}:
\begin{equation}
-{\rm Li}_{3/2}\left[-\xi\right]=\frac{4}{3\sqrt{\pi}}\Theta^{-3/2}.\label{eq:xi_Theta_relation}
\end{equation}
The conditions $\Gamma=1$ and $\Theta=1$ divide the density-temperature
parameter space into multiple regions, as seen in figure \ref{fig:parregimes}.
The regimes can be broken down into (1) classical weakly coupled,
(2) classical strongly coupled, (3) quantum weakly coupled, and (4)
classical strongly coupled ions with degenerate weak or strongly coupled
electrons. WDM exists at the intersection of all of these regions
marked by the blue region, where no small expansion parameter is available.
Transport in region (1) is well-understood in terms of the Landau-Spitzer
theory \citep{Spitzer1956}, and region (3) has been successfully
modeled through quantum weak-coupling theories such as the quantum
Landau-Fokker-Planck equation \citep{Daligault2016a}. Progress has
recently been made extending classical plasma transport theory into
region (2) for $\Gamma\lesssim20$ through use of mean force kinetic
theory (MFKT) \citep{Baalrud2013,Baalrud2014,Baalrud2019}, which
has also been successfully applied in region (4) for WDM in the case
of ion transport \citep{Daligault2016}. Other existing kinetic methods
for predicting transport in WDM typically fall into the categories
of binary collision theories \citep{Daligault2016a,Daligault2018,Gericke2002,Lee1984,Starrett2018},
linear response theories \citep{Daligault2009,Lampe1968,Scullard2018},
and non-equilibrium Green's functions and field-theoretic methods
\citep{Brown2005,Brown2007,Balzer2013,Bonitz2016}.

The model presented in this work is physically intuitive, contains
much of the relevant physics, and can be evaluated relatively quickly.
It is based on the Uehling-Uhlenbeck equation (named BUU equation
from this point on, with the letter B referencing Boltzmann), which
accounts for degeneracy and diffraction \citep{Uehling1933}. Correlations
in a moderate Coulomb coupling regime are modeled through the assertion
that the binary scattering is mediated by the equilibrium statistical
potential of mean force. This mean force is computed using a recent
combined Average-Atom + Two-Component-Plasma model \citep{Starrett2013,Starrett2018}.
The result has the advantage of retaining the dominant aspects of
the relevant physics, while remaining relatively fast to evaluate
in comparison to fully dynamical calculations. In the classical limit
the model can be rigorously derived \citep{Baalrud2019}, but while
this derivation cannot be easily extended to the quantum domain due
to the uncertainty principle, it is reasonable to apply the potential
of mean force to the BUU equation. 

Explicit results are computed for momentum and energy relaxation rates
of aluminum at conditions spanning the WDM regime. The results for
energy relaxation are found to be equivalent to a recent model by
Daligault and Simoni \citep{Daligault2019} if interactions are assumed
to occur via the potential of mean force in that theory. An unanticipated
result is observed for momentum relaxation, whereby degeneracy influences
the relaxation rate in a different manner than for energy relaxation.
This effect is not predicted by previous reduced kinetic theories,
but appears to lead to better agreement with quantum molecular dynamics
simulations of electrical conductivity at WDM conditions \citep{Witte2018}.

We begin by detailing the model in section \ref{sec:theory}. We introduce
the potential of mean force into the BUU equation and discuss what
the concept means in the context of a degenerate plasma. In section
\ref{sec:Relaxation} we apply this to electron-ion momentum and temperature
relaxation, where we obtain degeneracy- and correlation-dependent
``Coulomb integrals'' that replace the traditional Coulomb logarithm.
In section \ref{sec:discussion}, we evaluate the model for the solid-density
aluminum and compare to common and simple alternatives and discuss
the relative importance of the effects of correlation, large-angle
scattering, Pauli blocking, and diffraction. We conclude and summarize
in section \ref{sec:conclusions}.

\section{\label{sec:theory}A Kinetic Equation for Transport in WDM}

\subsection{The Uehling-Uhlenbeck Collision Operator}

We consider the collision integral from the right hand side of the
BUU equation \citep{Uehling1933},
\begin{gather}
C_{q}^{ss'}=\int d\boldsymbol{v}'d\Omega\frac{d\sigma}{d\Omega}u\biggl[\hat{f_{s}}\hat{f_{s'}}\left(1+\theta_{s}f_{s}\right)\left(1+\theta_{s'}f_{s'}\right)\nonumber \\
-f_{s}f_{s'}\left(1+\theta_{s}\hat{f_{s}}\right)\left(1+\theta_{s}\hat{f_{s'}}\right)\biggr]\label{eq:collisionoperator}
\end{gather}
where the ``hatted'' quantities $\hat{f_{s}}$ are evaluated at
the post-collision velocity $\hat{\boldsymbol{v}}=\boldsymbol{v}+\Delta\boldsymbol{v}$
and $\theta_{s}=(\pm1/g_{s})(h/m_{s})^{3}$ where $g_{s}$ is an integer
accounting for particle statistics with $g_{s}=g_{e}=2$ for electrons,
the $+$ sign corresponds with Bosons and the $-$ sign with Fermions.
Calculation of $d\sigma/d\Omega$ is carried out via a partial wave
expansion in terms of the phase shifts $\delta_{l}(\eta)$. The determination
of the phase shifts from the Schr\"{o}dinger equation is discussed
in Appendix \ref{sec:phaseshifts}.

The BUU equation describes the evolution of the Wigner quasi-probability
distribution function $f_{s}$. It was originally proposed as an extension
of the Boltzmann equation to account for degeneracy \citep{Uehling1933},
but a consistent derivation of the equation was not accomplished for
some time. Early methods involved applying the BBGKY hierarchy to
the kinetic equation for the Wigner function and often fell short
of fully obtaining the BUU equation, i.e. to include the $\theta_{s}$
terms \citep{Hoffman1965,Imam-Rahajoe1967}. Ultimately, a derivation
was carried out using the BBGKY hierarchy in the density operator
formalism \citep{Boercker1979}. This required a modification of the
typical weak-correlation assumption in derivations of the Boltzmann
equation. Instead of neglecting three-body correlations entirely,
Boercker and Dufty included the quantum correlations of two scattering
particles with a third spectator particle to preserve Fermion anti-symmetry,
without including correlations due to the interaction. By this method
they self-consistently derived the BUU equation with the statistical
$\theta_{s}$ factors, but came to the conclusion that the degeneracy
must simultaneously affect the scattering cross section in addition
to the statistical availability of scattering states encapsulated
in the $\theta_{s}$ terms.

The BUU equation as originally formulated is applicable to moderately
dense gases in which degeneracy is present but the amount of correlation
is small. In the case of WDM, the equation has several deficiencies.
First, it depends on the degree of degeneracy, which in turn depends
on the electron number density and therefore the average ionization
state of the system, which must be provided as an input. Second, in
a plasma it is well known that transport rates predicted by equation
(\ref{eq:collisionoperator}) diverge if the cross section is computed
using the Coulomb potential because the Coulomb force is of an infinite
range. This is typically resolved in an ad hoc manner by enforcing
a large distance limit on the impact parameter. Third, the derivation
of the BUU equation, while including correlations due to Fermi statistics,
does not allow for correlations due to the interaction and thus applies
only in the limit of weak coupling. The remainder of this section
describes how all three deficiencies can be addressed in a consistent
fashion in the WDM regime\@.

For a tenuous and hot (read classical and weakly coupled) plasma the
equilibrium ionization state is determined by the Saha equation \Citep{Saha1921}.
The divergence in the Coulomb logarithm is related to the neglect
of correlation: in plasmas the collective affect of the surrounding
plasma introduces Debye screening that limits the range of the interaction.
A recent approach called ``mean force kinetic theory'' has derived
a self-consistent approach to plasmas through a new expansion parameter
of the BBGKY hierarchy \citep{Baalrud2019}. In standard derivations
of the Boltzmann equation, the BBGKY hierarchy is truncated via neglecting
correlations involving three or more particles and making certain
assumptions about two-particle correlations. In mean-force kinetic
theory the BBGKY hierarchy is re-arranged in terms of an expansion
parameter that is the difference between the exact non-equilibrium
distribution function and the its equilibrium limit. The hierarchy
is then truncated by assuming this difference is negligible for reduced
distribution functions in three or more particle coordinates; i.e.
that the high order correlations take their equilibrium values. The
result is a collision integral identical in form to that of the Boltzmann
equation, but in which the scattering particles interact through the
potential of mean force. In addition, there is a term on the left
hand side of the kinetic equation that enforces the non-ideality of
the equilibrium limit in the equation of state. The result is capable
of describing transport in weak to moderately coupled plasmas ($\Gamma\lesssim20$)
based on the equilibrium structural properties of the plasma.

\subsection{The Quantum Potential of Mean Force}

Extending mean-force kinetic theory to include quantum effects is
complicated by two issues: the exclusion principle complicates the
mathematics of the necessary statistical averaging, and more significantly
the uncertainty principle muddles the very meaning of a potential
of mean force. Classically, the mean force is the force experienced
between two particles at rest with a given separation, with a statistical
averaging over all of the remaining particles in the plasma at equilibrium.
In the quantum case, knowing particles are ``at rest with a given
separation'' is impossible according to the uncertainty principle.
Mathematically, this prevents factoring of the kinetic and potential
(configuration) terms in the equilibrium density matrix, and ultimately
prevents a general derivation of the potential of mean force by extension
of known classical means.

Despite this complication, the potential of mean-force must have some
meaning in at least a semi-classical sense. An electron-ion pair will
still induce well-defined correlations in the plasma, and these correlations
can in turn influence the force felt by the interacting pair at least
over the average of many scattering events at many velocities. This
is reflected in the the screened potential 
\begin{equation}
U_{{\rm sc}}(r)=\frac{\phi(r)}{k_{{\rm B}}T}{\rm e}^{-r/\lambda_{{\rm sc}}}\label{eq:Uscreen}
\end{equation}
with degeneracy-dependent screening length (as per \citep{Melrose2010})
\begin{equation}
\lambda_{{\rm sc}}^{2}=\lambda_{{\rm D}}^{2}\sqrt{\frac{{\rm Li}_{3/2}\left(-\xi\right)}{{\rm Li}_{1/2}\left(-\xi\right)}}\label{eq:rsc}
\end{equation}
which can be seen as a weak-correlation limit of the potential of
mean force both for classical and quantum plasmas. The essential challenge
of applying the mean force concept to WDM is how to encapsulate this
effect in a binary potential when the coupling is no longer weak.
It has long been known that weak correlations influence the potential
in the form of plasma screening in both the classical (Debye-Huckel)
and quantum (Thomas-Fermi limits). One other classical derivation
of the potential of mean force is via the Ornstein-Zernike equation,
which defines the direct correlation function \citep{Hansen2006}.
Fortunately, a quantum analog of the Ornstein-Zernike equation exists,
and this equation has been used successfully to calculate the equilibrium
pair correlation function in WDM \citep{Starrett2012,Starrett2013,Starrett2018}.
Furthermore, it has been used to define a quantum potential of mean
force for electron-ion interactions, and this potential has been used
to predict electrical conductivities with good agreement with quantum
molecular dynamics simulations \citep{Shaffer2020}.
\begin{figure}[b]
\includegraphics[width=8.55cm]{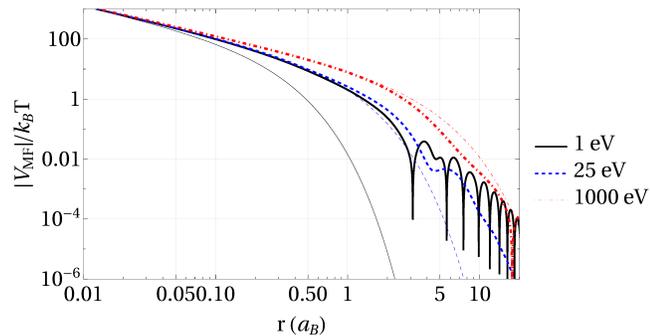}\caption{\label{fig:gii_Vie} Electron-ion potential of mean force for solid-density
($2.7\thinspace{\rm g}\cdot{\rm cm}^{-3}$) warm dense aluminum at
$1$, $25$ and $500$ eV (thick curves). In the high-temperature
limit the potential approaches the screened Coulomb potential of a
classical plasma (thin curves); as the temperature decreases it is
altered by both degeneracy and correlations leading to different scale
lengths of the potential in addition to the non-monotonic behavior.}
\end{figure}

We postulate that a quantum potential of mean force must arise naturally
from the quantum mean force kinetic theory, and that this potential
is that which is derived from the quantum Ornstein-Zernike equation.
We turn to such a potential obtained with the quantum hypernetted-chain-approximation,
coupled with an average-atom model that accounts for the structure
and ionization state of the ions \citep{Starrett2012,Starrett2013,Starrett2018,Starrett2017}
(subsequently referred to as the AA-TCP model for ``Average-Atom
Two-Component-Plasma''. This potential can be expressed as
\begin{gather}
V^{{\rm MF}}(r)=-\frac{Z}{r}+\int d^{3}r'\frac{n_{e}^{{\rm ion}}(r')}{|\boldsymbol{r}-\boldsymbol{r'}|}\nonumber \\
+V^{{\rm xc}}[n_{e}^{{\rm ion}}(r)]+n_{i}^{0}\int d^{3}r'\frac{C_{ie}(|\boldsymbol{r}-\boldsymbol{r'}|)}{-\beta}h_{ii}(r')\nonumber \\
+\bar{n}_{e}^{0}\int d^{3}r'\frac{C_{ee}(|\boldsymbol{r}-\boldsymbol{r'}|)}{-\beta}h_{ie}(r'),
\end{gather}
where $C_{ie}$ and $C_{ee}$ are the electron-ion and electron-electron
direct correlation functions respectively, $h_{ie}$ and $h_{ee}$
are the electron-ion and electron-electron pair correlation functions
respectively, $n_{e}^{{\rm ion}}(r)$ is the density of bound electrons,
$\beta=k_{B}T$, and $V^{{\rm xc}}$ is the exchange correlation functional
(in the case of \citep{Starrett2013} it is the zero-temperature Dirac
exchange functional \citep{Dirac1930}). Calculation of the potential
requires closure, which in this case is provided by the quantum hypernetted-chain-approximation
for the ion-ion correlations and through coupling to an Average-Atom
model for the electron-ion correlations. Such methods can be substantially
faster than full dynamical calculations such as molecular dynamics,
wherein lies the primary benefit of the theory proposed in this work.
In figure \ref{fig:gii_Vie} we show example electron-ion scattering
potentials from the AA-TCP model for warm dense aluminum at conditions
that span the weakly coupled classical to moderately coupled degenerate
regimes; see regions (1) and (4) of figure \ref{fig:parregimes}.
The figure demonstrates the convergence of the potential of mean force
with a screened Coulomb potential in the weakly-coupled limit, and
the importance of correlations in the calculation of the potential
in the region of moderate coupling.

\section{Transport Rates\label{sec:Relaxation}}

Comprehensive methods to derive hydrodynamic equations, such as that
of Chapman and Enskog have been developed for the Boltzmann equation
\citep{ChapmanCowling}, but their extension to the BUU equation faces
considerable mathematical challenges and has not been accomplished
to our knowledge. To demonstrate predictions for macroscopic transport
rates, we focus on electron-ion relaxation in which the respective
electron and ion distribution functions are known but the species
are not in equilibrium with each other. We consider both temperature
relaxation and momentum relaxation, which is related to the electrical
conductivity. A restriction imposed by considering only electron-ion
relaxation is that it provides only one contribution to processes
such as electrical conductivity that are also influenced by electron-electron
interactions. Although models such as the quantum Landau-Fokker-Planck
equation have been solved using a Chapman-Enskog technique to address
both contributions in a comprehensive hydrodynamic theory \citep{Daligault2018},
they do not address strong coupling. A recent modification has been
proposed to incorporate strong coupling via a modified Coulomb logarithm
computed using the potential of mean force, and finds that in the
strongly degenerate regime and for high-Z systems the electron-ion
collisions are dominant \citep{Shaffer2020}. However, the Fokker-Planck
form of the collision operator itself is only expected to apply when
momentum transfer during collisions is small (i.e., weak coupling).
For instance, it can be derived from a small momentum transfer expansion
of the BUU equation. Here, we focus on the electron-ion relaxation
using the full BUU equation in order to isolate the influence of large
momentum transfer in the collision operator. 

Concentrating on the electron-ion contribution also allows for a commensurable
comparison with quantum MD simulations of electrical conductivity
\citep{Witte2018}. Since electrons are often treated using the Born-Oppenhiemer
approximation in simulations, they are also limited to treat only
the electron-ion contribution to transport processes. Although electron-electron
interactions are expected to contribute to the total conductivity,
it is only recently becoming possible to simulate dynamic electrons
in WDM following advancements in wave-packet MD \citep{Ma2019}, mixed
quantum-classical MD \citep{Simoni2019,Daligault2018a}, Bohmian quantum
methods \citep{Larder2019}, Kohn-Sham DFT \citep{White2020} and
quantum Monte Carlo \citep{Yilmaz2020}. Addressing contributions
from both electron and ion dynamics will be the next step in both
the theory and simulation development.

\subsection{General Formalism}

A binary mixture of two species $s$ and $s'$ out of equilibrium
will relax towards equilibrium through $s-s$, $s-s'$ and $s'-s'$
collisions, which are modeled by moments of the collision operator
(\ref{eq:collisionoperator}),
\begin{equation}
\left\langle \chi\right\rangle ^{s-s'}=\int d\boldsymbol{v}\chi\left(\boldsymbol{v}\right)C_{qB}^{s-s'},
\end{equation}
where $\chi(\boldsymbol{v})$ is some polynomial function of the velocity.
To simplify, we utilize the following properties: $d\Omega\frac{d\sigma}{d\Omega}$
is invariant under reversal of the collision, i.e. $\left(\boldsymbol{\boldsymbol{v}},\boldsymbol{v}'\right)\leftrightarrow\left(\hat{\boldsymbol{v}},\hat{\boldsymbol{v}}'\right)$
where $\boldsymbol{\boldsymbol{v}}\thinspace{\rm and\thinspace}\boldsymbol{v}'$
are the pre-collision velocities of particles one and two respectively,
the ``hat'' $\hat{}$ indicates a post-collision quantity, and the
phase-space volume element is invariant, i.e. $\int d\boldsymbol{v}d\boldsymbol{v}'=\int d\hat{\boldsymbol{v}}d\hat{\boldsymbol{v}'}$.
We thus obtain
\begin{gather}
\left\langle \chi\right\rangle ^{s-s'}=\int d\boldsymbol{v}\int d\Omega\frac{d\sigma}{d\Omega}u\int d\boldsymbol{v}'\left[\chi\left(\hat{\boldsymbol{v}}\right)-\chi\left(\boldsymbol{v}\right)\right]\nonumber \\
\times f_{s}f_{s'}\left(1+\theta_{s}\hat{f_{s}}\right)\left(1+\theta_{s}\hat{f_{s'}}\right).\label{eq:chimoments}
\end{gather}
Relevant $\chi\left(v\right)$ include

\begin{equation}
\chi\left(\boldsymbol{v}\right)=\begin{cases}
1\thinspace\thinspace\thinspace\thinspace\thinspace\thinspace\thinspace\thinspace\thinspace\rightarrow & \left[\chi\left(\hat{\boldsymbol{v}}\right)-\chi\left(\boldsymbol{v}\right)\right]=0\\
m_{s}\boldsymbol{v}\thinspace\rightarrow & \left[\chi\left(\hat{\boldsymbol{v}}\right)-\chi\left(\boldsymbol{v}\right)\right]=m_{s}\boldsymbol{\Delta}v\\
m_{s}v^{2}\rightarrow & \left[\chi\left(\hat{\boldsymbol{v}}\right)-\chi\left(\boldsymbol{v}\right)\right]=m_{s}\Delta v^{2}
\end{cases}
\end{equation}
where $\Delta\boldsymbol{v}=\hat{\boldsymbol{v}}-\boldsymbol{v}$.
Substituting variables $\boldsymbol{v}=\boldsymbol{v}'+\boldsymbol{u}$,
defining $m_{ss'}=m_{s}m_{s'}/(m_{s}+m_{s'}),$ and utilizing the
following relations obtained from the collision kinematics: $m_{s}\Delta\boldsymbol{v}=m_{ss'}\Delta\boldsymbol{u},$$\Delta\boldsymbol{u}\cdot\Delta\boldsymbol{u}=-2\boldsymbol{u}\cdot\Delta\boldsymbol{u}$and
$\left(2\boldsymbol{v}\cdot\Delta\boldsymbol{v}+\Delta\boldsymbol{v}^{2}\right)=(m_{ss'}/m_{s})\Delta\boldsymbol{u}\cdot\left[\boldsymbol{v}'+(m_{ss'}/m_{s})\boldsymbol{u}\right],$
shows that (see \citep{Baalrud2012}),
\begin{gather}
\chi\left(\boldsymbol{u}\right)=\nonumber \\
\begin{cases}
1\thinspace\thinspace\thinspace\thinspace\thinspace\thinspace\thinspace\thinspace\thinspace\rightarrow & \left[\chi\left(\hat{\boldsymbol{v}}\right)-\chi\left(\boldsymbol{v}\right)\right]=0\\
m_{s}\boldsymbol{v}\thinspace\rightarrow & \left[\chi\left(\hat{\boldsymbol{v}}\right)-\chi\left(\boldsymbol{v}\right)\right]=m_{ss'}\Delta\boldsymbol{u}\\
m_{s}v^{2}\rightarrow & \left[\chi\left(\hat{\boldsymbol{v}}\right)-\chi\left(\boldsymbol{v}\right)\right]=m_{ss'}\left(\boldsymbol{v'}-\boldsymbol{V}_{s}+\frac{m_{ss'}}{m_{s'}}\boldsymbol{u}\right)\cdot\Delta\boldsymbol{u}
\end{cases}
\end{gather}
where 
\begin{equation}
\Delta\boldsymbol{u}=u\left({\rm sin}\theta{\rm cos}\phi\hat{\boldsymbol{x}}+{\rm sin}\theta{\rm sin}\phi\hat{\boldsymbol{y}}-2{\rm sin}^{2}\frac{\theta}{2}\hat{\boldsymbol{u}}\right).
\end{equation}

The preceding discussion and the collision operator (\ref{eq:collisionoperator})
are in principle applicable to transport in any semi-classical system.
As it pertains to WDM, ion-ion scattering is contained within this
formalism as ion dynamics are classical and electron degeneracy effects
enter only via the potential of mean force. Application of the theory
to ion-ion scattering was validated in \citep{Daligault2016}. The
case of the electron-electron terms requires further work due to the
subtleties associated with defining the potential of mean force that
are discussed in section \ref{sec:theory} and will be investigated
in another work. However, the model at the level to which we have
developed it has immediate applicability to the case of electron-ion
scattering.

\subsection{The Relaxation Problem}

We restrict our analysis to the class of problems in which electrons
and ions in the plasma are in respective equilibrium with themselves
with different fluid quantities $T_{e},\thinspace T_{i},\thinspace\boldsymbol{V}_{e}\thinspace{\rm and}\thinspace\boldsymbol{V}_{i}$,
respectively. In such a system, the electron and ion fluid variables
will equilibrate on a timescale long compared to the respective electron-electron
and ion-ion collision times. The ions have a classical Maxwellian
velocity distribution
\begin{equation}
f_{i}\left(\boldsymbol{v}'\right)=\frac{n_{i}}{v_{Ti}^{3}}\frac{e^{-\left(\boldsymbol{v}'-\boldsymbol{V}_{i}\right)^{2}/v_{Ti}^{2}}}{\pi^{3/2}}
\end{equation}
and the electrons have a Fermi-Dirac velocity distribution
\begin{equation}
f_{e}\left(\boldsymbol{v}\right)=n_{e}\left[v_{Te}^{3}\left(-\pi^{3/2}\text{Li}_{\frac{3}{2}}(-\xi)\right)\left(1+\frac{{\rm e}^{\left(\boldsymbol{v}-\boldsymbol{V}_{e}\right)^{2}/v_{Te}^{2}}}{\xi}\right)\right]^{-1}
\end{equation}
where $v_{Ts}=\sqrt{2k_{B}T_{s}/m_{s}}$ and $\xi=\exp\left(\mu/k_{B}T\right)$,
the ion velocity is $\boldsymbol{v'}$ and electron velocity is $\boldsymbol{v}$.
We can write
\begin{gather}
f_{e}f_{i}\left(1+\theta_{e}\hat{f_{e}}\right)=\frac{n_{i}}{v_{Ti}^{3}}\frac{e^{-\left(\boldsymbol{v}'-\boldsymbol{V}_{i}\right)^{2}/v_{Ti}^{2}}}{\pi^{3/2}}n_{e}\nonumber \\
\times\left[v_{Te}^{3}\left(-\pi^{3/2}\text{Li}_{\frac{3}{2}}(-\xi)\right)\left(1+\frac{{\rm e}^{\left(\boldsymbol{v}'+\boldsymbol{u}-\boldsymbol{V}_{e}\right)^{2}/v_{Te}^{2}}}{\xi}\right)\right]^{-1}\nonumber \\
\times\left[1-\left(1+\frac{{\rm e}^{\left(\boldsymbol{v}'+\boldsymbol{u}+(m_{ei}/m_{e})\Delta\boldsymbol{u}-\boldsymbol{V}_{e}\right)^{2}/v_{Te}^{2}}}{\xi}\right)^{-1}\right],\label{eq:integrand1}
\end{gather}
from which the relation of the factor $\left(1+\theta_{e}\hat{f_{e}}\right)$
to Pauli blocking can be seen in terms of the Fermi-Dirac occupation
number: the contribution to the collision integral from collisions
to or from occupied states is zero. This simplification occurs from
the combination of $\theta_{e}=(-1/2)(h/m_{s})^{3}$ with the prefactor
$n_{e}v_{Te}^{3}/\text{Li}_{\frac{3}{2}}(-\xi)$ in the Fermi Dirac
distribution through the relation (\ref{eq:xi_Theta_relation}). 

Electron-ion temperature and momentum relaxation rates depend on the
energy exchange density $Q^{s-s'}$ and friction force density $\boldsymbol{R}^{s-s'}$,
respectively. These can in turn be written in terms of the moments
(\ref{eq:chimoments}), assuming a uniform plasma, as 
\begin{equation}
Q^{ei}=\left\langle \frac{1}{2}m_{e}\left(\boldsymbol{v}-\boldsymbol{V}_{e}\right)^{2}\right\rangle ^{e-i}=\frac{3n_{e}}{2}\frac{dT_{e}}{dt}\label{eq:Qei}
\end{equation}
(where in the last equality we have taken $\boldsymbol{V}_{e}=\boldsymbol{V}_{i}=0)$
and 
\begin{equation}
\boldsymbol{R}^{ei}=\left\langle m_{e}\boldsymbol{v}\right\rangle ^{e-i}=m_{e}\frac{d\boldsymbol{V}_{e}}{dt}\label{eq:Rei}
\end{equation}
which, in the respective limits of $\Delta T\ll T$ and $\Delta V\ll V$
yield simple relaxation rates $dT_{e}/dt=\nu_{ei}^{(\epsilon)}\Delta T$
and $d\boldsymbol{V}_{e}/dt=\nu_{ei}^{(p)}\Delta\boldsymbol{V}$.

The integration over the ion velocity can be simplified significantly
in the limit that the ion velocities are much smaller than the electron
velocities: $m_{e}T_{i}\ll m_{i}T_{e}$, which (due to the small electron-to-ion
mass ratio) is true when temperature differences are not extreme,
coinciding with our expansion about the equilibrium state. Note that
we also make the simplifying replacement $m_{ei}\approx m_{e}$. By
expanding equation (\ref{eq:integrand1}) in the limit that the electron
distribution is approximately constant over the range of accessible
ion velocities, the integral over the ion velocities can be carried
out analytically. The evaluation of this integral differs for the
calculation of $Q^{ei}$ versus $\boldsymbol{R}^{ei}$. Therefore
we examine each case separately.

\subsubsection{Temperature Relaxation}

\begin{figure}[!tph]
\includegraphics[width=8.55cm]{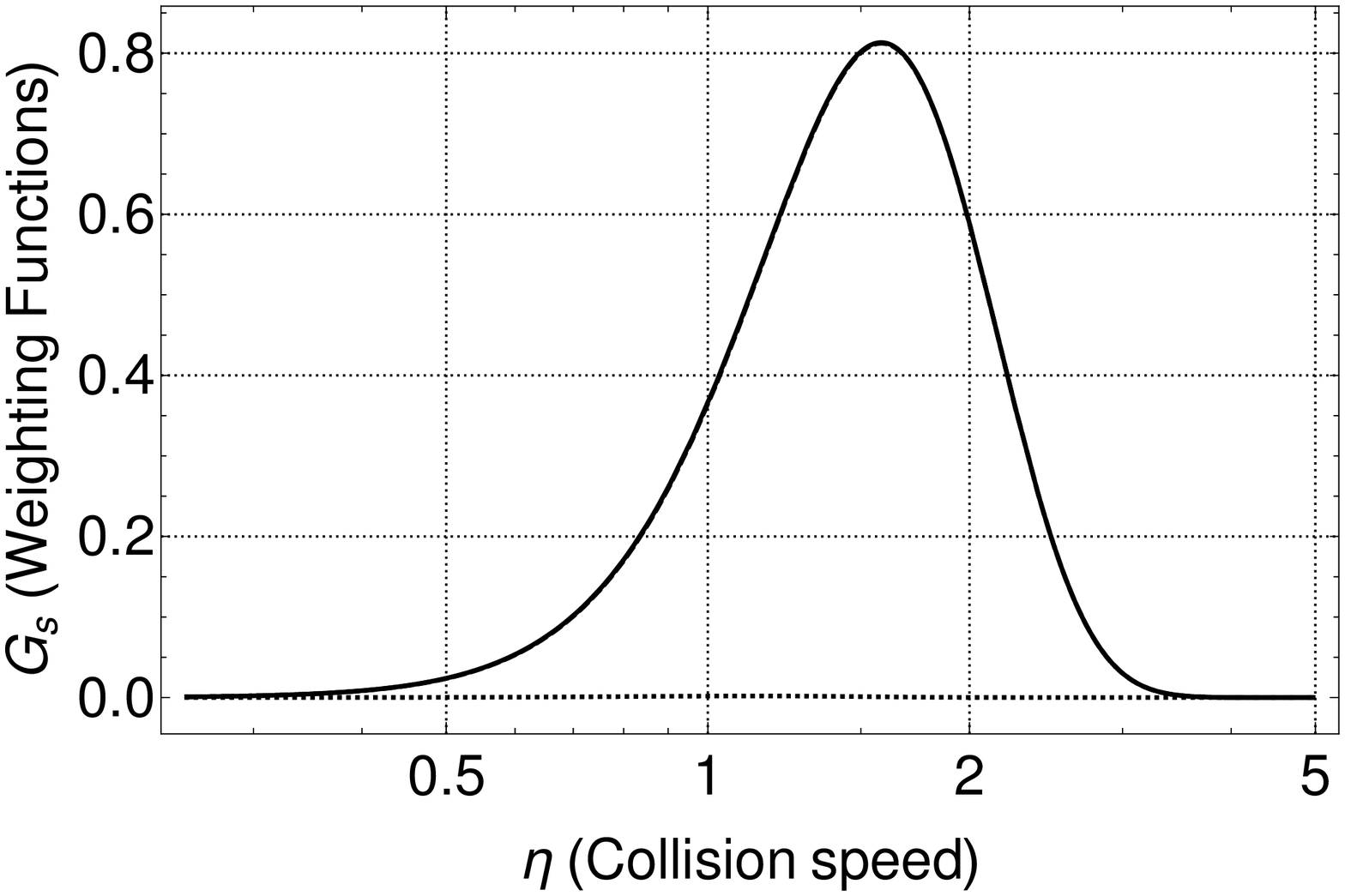}\\
\includegraphics[width=8.55cm]{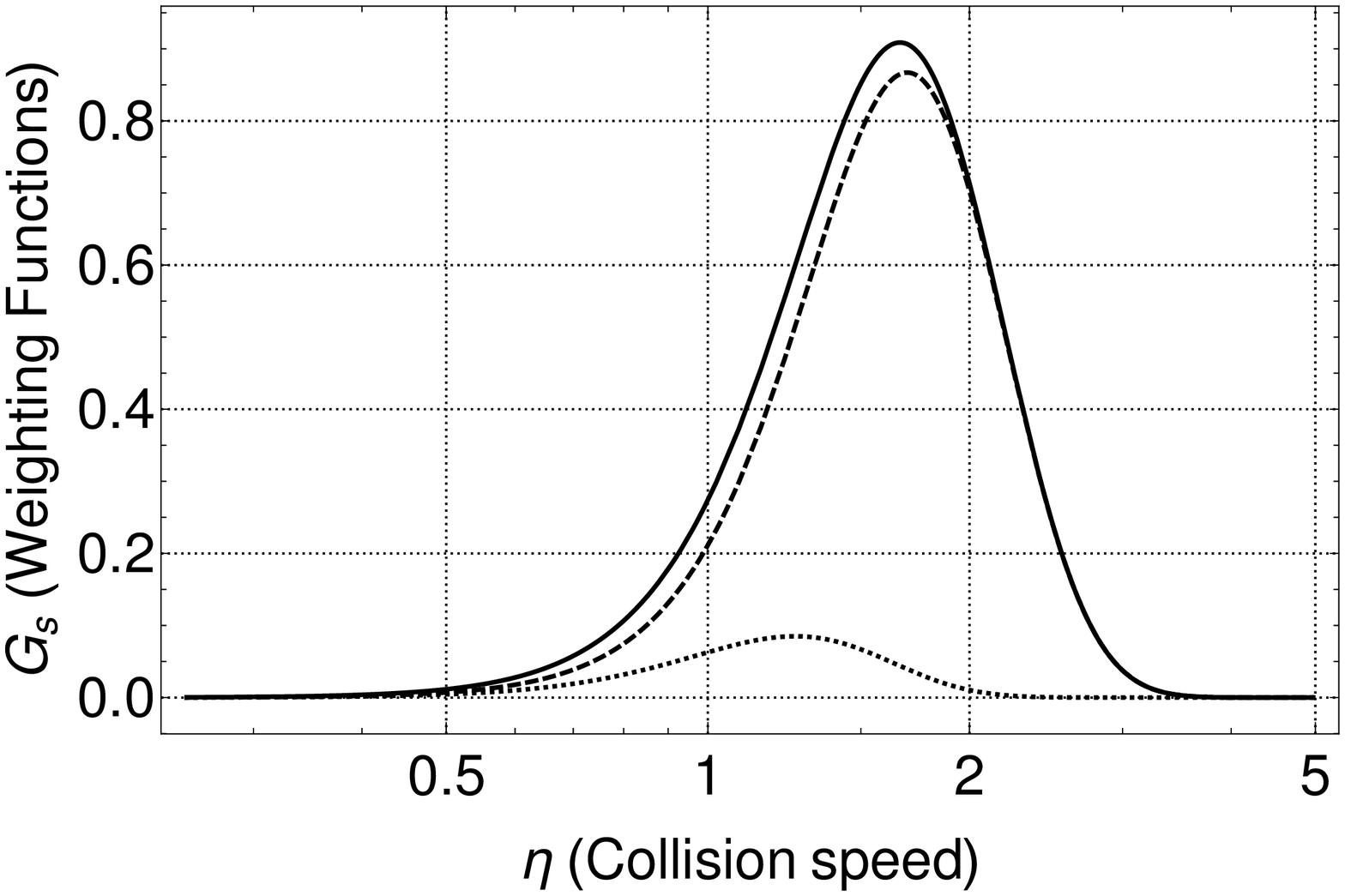}\\
\includegraphics[width=8.55cm]{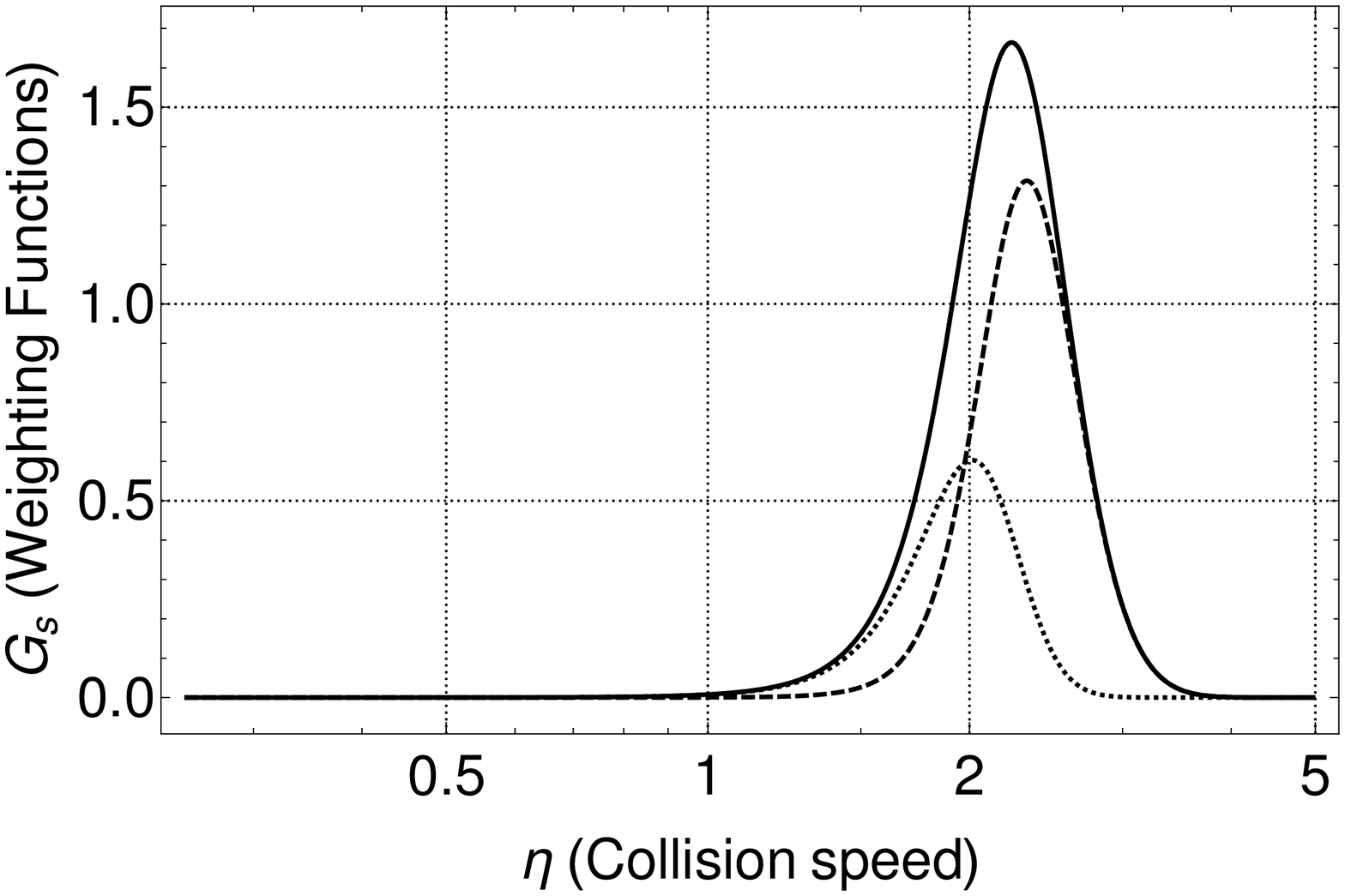}

\caption{\label{fig:integrands} Statistical contributions to the integrands
for temperature and momentum relaxation, $G_{1}$ (solid), $G_{2}$
(dashed) and $G_{3}$ (dotted), for three conditions: $\Theta=12.6$
and $\xi=0.017$ (top, weak degeneracy), $\Theta=0.85$ and $\xi=1.33$
(middle, moderate degeneracy), $\Theta=0.14$ and $\xi=1135$ (bottom,
strong degeneracy). The relevant collision velocities for both momentum
and temperature relaxation become narrowly centered around the Fermi
velocity at strong degeneracy. The relative importance of the two
different functions that contribute to momentum relaxation is degeneracy
dependent.}
\end{figure}

The energy-exchange density (\ref{eq:Qei}) in this case becomes 
\begin{gather*}
Q^{ei}=\int d\boldsymbol{u}\int d\Omega\frac{d\sigma}{d\Omega}u\boldsymbol{\Delta u}\\
\cdot\int d\boldsymbol{v}'m_{ei}\left(\boldsymbol{v'}+\frac{m_{ei}}{m_{i}}\boldsymbol{u}\right)f_{i}f_{e}\left(1-|\theta_{e}|\hat{f_{e}}\right).
\end{gather*}
Inserting equation (\ref{eq:integrand1}), applying the expansion
$|\boldsymbol{v}'|\ll|\boldsymbol{u}|$, assuming zero drift velocities
and $\left|T_{e}-T_{i}\right|\ll T_{e},T_{i}$ we perform the integral
over $\boldsymbol{v}'$ and write
\begin{gather*}
\boldsymbol{\Delta u}\cdot\int d\boldsymbol{v}'m_{ei}\left(\boldsymbol{v'}+\frac{m_{ei}}{m_{i}}\boldsymbol{u}\right)f_{i}f_{e}\left(1-|\theta_{e}|\hat{f_{e}}\right)\approx\\
\frac{n_{e}n_{i}\eta\xi e^{-\eta^{2}}\sin^{2}\left(\frac{\theta}{2}\right)}{\pi^{3}v_{Te}\text{Li}_{\frac{3}{2}}\left(-\xi\right)\left(\xi e^{-\eta^{2}}+1\right)^{2}}
\end{gather*}
where $\eta\equiv u/v_{Te}$. The result is written to facilitate
comparison with the classical limit,
\begin{equation}
Q^{ei}=-3\frac{m_{e}}{m_{i}}n_{e}\nu_{ei}^{(\epsilon)}(T_{e}-T_{i}).\label{eq:Trelax}
\end{equation}
in terms of a collision frequency
\begin{equation}
\nu_{ei}^{(\epsilon)}=\nu_{0}\Xi_{ei}^{(\epsilon)},\label{eq:collfrequencyenergy}
\end{equation}
where
\begin{equation}
\nu_{0}\equiv\frac{4\sqrt{2\pi}n_{i}Z^{2}e^{4}}{3\sqrt{m_{e}}(k_{B}T_{e})^{3/2}}=2.906\times10^{-12}\frac{Zn_{i}[{\rm m}^{-3}]}{(T_{e}[{\rm eV}])^{3/2}}
\end{equation}
 and a generalized Coulomb integral $\Xi_{ei}^{(\epsilon)}$. Effects
of degeneracy and strong coupling are contained in the Coulomb integral,
\begin{gather}
\Xi_{ei}^{(\epsilon)}=\frac{1}{2}\int_{0}^{\infty}d\eta I_{\epsilon}(\eta)\label{eq:xiE1}\\
I_{\epsilon}(\eta)\equiv G_{1}(\eta)\frac{\sigma_{1}^{(1)}\left(\eta,\Gamma\right)}{\sigma_{0}}\nonumber 
\end{gather}
where
\begin{equation}
\sigma_{1}^{(1)}\left(\eta,\Gamma\right)=4\pi\int_{0}^{\pi}d\theta\sin^{2}\frac{\theta}{2}\sin\theta\frac{d\sigma}{d\Omega}
\end{equation}
is the momentum transfer cross section, which can be written in terms
of the phase shifts $\delta_{l}$ as
\begin{equation}
\sigma_{1}^{(1)}=\frac{4\pi}{\eta^{2}}\sum_{l=0}^{\infty}(l+1)\sin^{2}(\delta_{l+1}-\delta_{l}).\label{eq:momCC1}
\end{equation}
 The function
\begin{equation}
G_{1}(\eta)\equiv\frac{\xi{\rm e}^{-\eta^{2}}\eta^{5}}{\left[-\text{Li}_{\frac{3}{2}}\left(-\xi\right)\right]\left(\xi e^{-\eta^{2}}+1\right)^{2}}
\end{equation}
determines the relative availability of states that contribute to
the scattering. This is plotted in figure \ref{fig:integrands} for
several values of the degeneracy parameter $\Theta$, where it is
shown that in the classical limit scattering is dominated by energy
transfers around the thermal energy, and as degeneracy increases the
envelope of relevant energy-transfers narrows about the Fermi energy.
It should be noted that the relaxation rate obtained in equation (\ref{eq:collfrequencyenergy})
is identical to that obtained in equation (71) of reference \citep{Daligault2019}
by very different means.

\subsubsection{Momentum Relaxation}

Momentum relaxation occurs through collisions between electron and
ion populations with different average velocities. The force density
(\ref{eq:Rei}) associated with these collisions is

\begin{equation}
\boldsymbol{R}^{ei}=\int d\boldsymbol{u}\int d\Omega\frac{d\sigma}{d\Omega}u\int d\boldsymbol{v}'m_{ei}\boldsymbol{\Delta u}f_{e}f_{i}\left(1+\theta_{e}\hat{f_{e}}\right).
\end{equation}
Inserting equation (\ref{eq:integrand1}), applying the expansion
$|\boldsymbol{v}'|\ll|\boldsymbol{u}|,$ and assuming $\left|T_{e}-T_{i}\right|\ll T_{e},T_{i}$,
$V_{i}\ll v_{Ti}$ and $V_{e}\ll v_{Te}$, the integral over $\boldsymbol{v}'$
can be performed analytically,
\begin{gather*}
\int d\boldsymbol{v}'m_{ei}f_{e}f_{i}\left(1+\theta_{e}\hat{f_{e}}\right)\\
\approx\frac{2m_{e}n_{e}\xi n_{i}e^{-\eta^{2}}\left[\boldsymbol{u}\cdot\boldsymbol{\Delta V}-\xi e^{-\eta^{2}}\left(\boldsymbol{\Delta u}+\boldsymbol{u}\right)\cdot\boldsymbol{\Delta V}\right]}{\pi^{3/2}v_{Te}^{5}\left[-\text{Li}_{\frac{3}{2}}\left(-\xi\right)\right]\left(\xi e^{-\eta^{2}}+1\right){}^{3}}.
\end{gather*}
We follow the classical example and write
\begin{equation}
\boldsymbol{R}^{ei}=-n_{e}m_{e}\nu_{ei}^{(p)}(\boldsymbol{V}_{e}-\boldsymbol{V}_{i})\label{eq:momrelax}
\end{equation}
where the frequency
\begin{equation}
\nu_{ei}^{(p)}=\nu_{0}\Xi_{ei}^{(p)}\label{eq:nueip}
\end{equation}
involves a Coulomb integral
\begin{gather}
\Xi_{ei}^{(p)}=\frac{1}{2}\int_{0}^{\infty}d\eta I_{p}(\eta,\Gamma,\xi)\nonumber \\
I_{p}(\eta,\Gamma,\xi)\equiv G_{2}(\eta,\xi)\frac{\sigma_{1}^{(1)}\left(\eta,\Gamma\right)}{\sigma_{0}}-G_{3}(\eta,\xi)\frac{\sigma_{2}^{(1)}\left(\eta,\Gamma\right)}{\sigma_{0}}\label{eq:xiMOM1}\\
=G_{2}(\eta,\xi)\frac{\sigma_{p}(\eta,\Gamma,\xi)}{\sigma_{0}}
\end{gather}
which is different from that involved in the energy-exchange density.
Here, $\sigma_{1}^{(1)}$ is defined in equation (\ref{eq:momCC1}),
\begin{gather}
\sigma_{2}^{(1)}\left(\eta,\Gamma\right)=4\pi\int_{0}^{\pi}d\theta\sin^{2}\frac{\theta}{2}\sin\theta\frac{d\sigma}{d\Omega}\cos\theta,\label{eq:momCC2a}
\end{gather}
and
\begin{equation}
\sigma_{p}(\eta,\Gamma,\xi)=\sigma_{1}^{(1)}\left(\eta,\Gamma\right)-\xi{\rm e}^{-\eta^{2}}\sigma_{2}^{(1)}\left(\eta,\Gamma\right).\label{eq:sigmap}
\end{equation}
It is interesting to note that the cross section arising in the energy
relaxation rate from equation (\ref{eq:momCC1}) differs from that
associated with momentum relaxation from equation (\ref{eq:sigmap}).
This is a purely quantum mechanical effect, as the cross section definitions
are the same in the classical limit \citep{Baalrud2012}. It is also
an effect that is predicted by the full BUU equation, but not the
Landau-Fokker-Plank limit associated with small momentum transfer
interactions \citep{Shaffer2020}. The weighting functions 
\begin{gather}
G_{2}(\eta,\xi)=\frac{\xi{\rm e}^{-\eta^{2}}\eta^{5}}{\left[-\text{Li}_{\frac{3}{2}}\left(-\xi\right)\right]\left(\xi e^{-\eta^{2}}+1\right)^{3}},\\
G_{3}(\eta,\xi)=\frac{\xi^{2}{\rm e}^{-2\eta^{2}}\eta^{5}}{\left[-\text{Li}_{\frac{3}{2}}\left(-\xi\right)\right]\left(\xi e^{-\eta^{2}}+1\right)^{3}},
\end{gather}
are shown in figure \ref{fig:integrands} and compared with the statistical
weighting factors in the case of temperature relaxation. The presence
of the differing angular integrals between the energy and momentum
relaxation cases warrants further discussion.

Through the use of the Wigner-3j function, $\sigma_{2}^{(1)}$ can
be expanded in the phase shifts (see appendix \ref{sec:shifts_appendix})
as
\begin{widetext}
\begin{gather}
\sigma_{2}^{(1)}=\frac{4\pi}{\eta^{2}}\sum_{l=0}^{\infty}\frac{\sin\delta_{l}}{4l(l+1)-3}\nonumber \\
\times\left\{ (l+1)(2l-1)\left[(l+2)\sin(\delta_{l}-2\delta_{l+2})-(2l+3)\sin(\text{\ensuremath{\delta_{l}}}-2\text{\ensuremath{\delta_{l+1}}})\right]-l^{2}(2l+3)\sin\delta_{l}\right\} .\label{eq:momCC2b}
\end{gather}
While it is tempting to interpret the quantity $\sigma_{2}^{(1)}$
as a cross-section, $\sigma_{2}^{(1)}$ can become negative and therefore
has no such interpretation. We will show in the next section that
it is only in the combination defined in equation (\ref{eq:sigmap})
that this interpretation is justified. We thus refer to $\sigma_{p}$
as an effective transport cross section. We note that this second
term arises due to degeneracy, and has no analog in the classical
relaxation problem.
\end{widetext}

\subsubsection{Electrical Conductivity}

The electrical conductivity is an important transport coefficient
that depends largely on the electron-ion collisional momentum relaxation
rate. Considering a Fermi-Dirac electron population flowing through
a stationary Maxwellian ion population due to an applied electric
field, the frictional force balances the electric force
\begin{equation}
\boldsymbol{R}_{ei}=-en_{e}\boldsymbol{E},
\end{equation}
which in the form of equation (\ref{eq:momrelax}) is connected to
the current through Ohm's law
\begin{equation}
\boldsymbol{J}=\sigma\boldsymbol{E}
\end{equation}
where $\boldsymbol{J}=-en_{e}\boldsymbol{V}_{e}$. Using the electron-ion
collisional friction {[}equation (\ref{eq:momrelax}){]}, the resulting
electrical conductivity is
\begin{equation}
\sigma=\frac{e^{2}n_{e}}{m_{e}\nu_{ei}^{(p)}},\label{eq:conductivity}
\end{equation}
where $\nu_{ei}^{(p)}$ is defined in equation (\ref{eq:nueip}).
The assumption of a Fermi-Dirac electron distribution means that electron-electron
(e-e) collisions do not contribute to the relaxation; distortions
in the electron distribution away from equilibrium amount to a higher
order approximation that could be explored e.g. through the Chapman-Enskog
expansion. The e-e collisions do not contribute substantially in the
degenerate regimes due to Pauli blocking, and at high temperatures
the e-e contribution is well understood via the Landau-Spitzer theory.
The intermediate regime where both degeneracy and e-e collisions are
important is discussed by Shaffer and Starrett \citep{Shaffer2020}
in the context of the quantum Fokker-Planck equation. The application
of the BUU equation to this regime to relax the assumption of small-momentum-transfer
collisions will require a Chapman-Enskog expansion of the BUU equation
and will be addressed in further studies.

\section{\label{sec:discussion}Results and Discussion}

\begin{figure}[!tph]
\includegraphics[width=8.55cm]{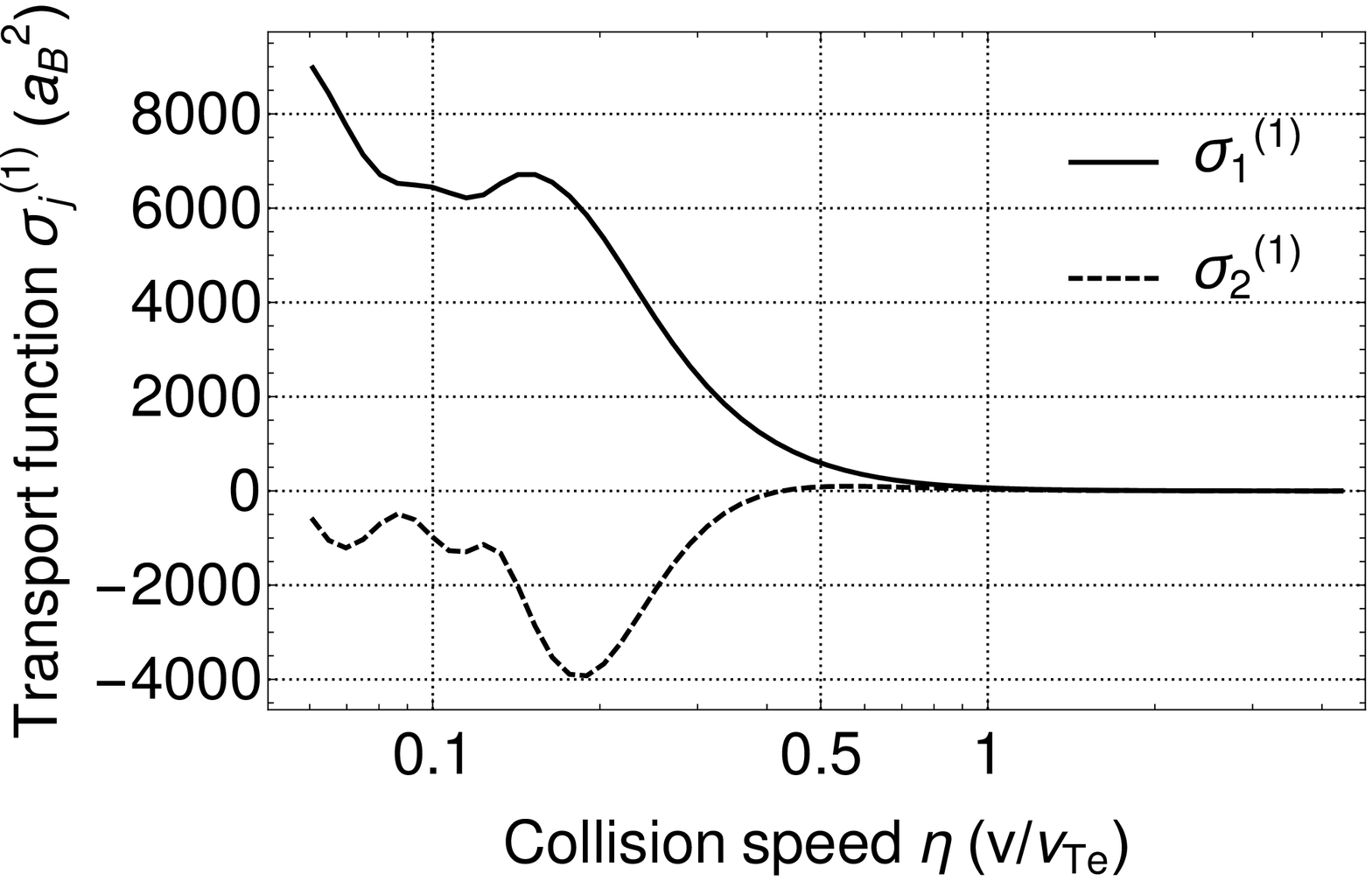}\\
\includegraphics[width=8.55cm]{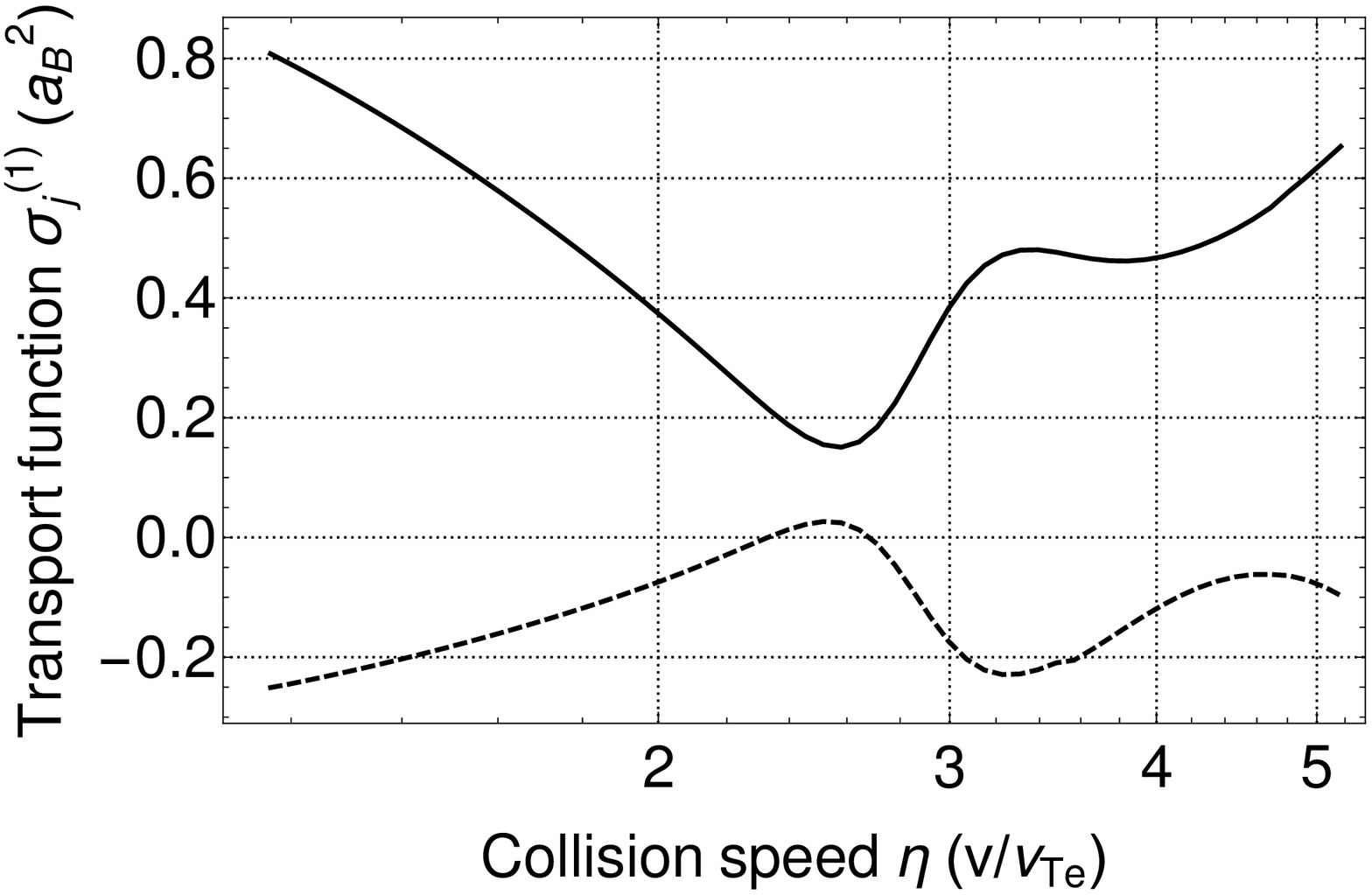}\caption{\label{fig:momCC} Functions $\sigma_{1}^{(1)}$ and $\sigma_{2}^{(1)}$
calculated using the PMF for solid density $(2.7\thinspace{\rm g}\cdot{\rm cm}^{-3})$
aluminum at $1000\thinspace{\rm eV}$ (top) and $1\thinspace{\rm eV}$
(bottom). While $\sigma_{2}^{(1)}$ plays the role of a cross section,
it is evident from its negative value at many velocities that it is
not one. Interestingly, it behaves (only approximately) inversely
to the momentum transfer cross section $\sigma_{1}^{(1)}$. While
$\sigma_{2}^{(1)}$ is non-zero at high temperatures, its influence
is negligible due to the suppression of the term that contains it
when $T\gg E_{{\rm F}}$.}
\end{figure}

\begin{figure}[!tph]
\includegraphics[width=8.55cm]{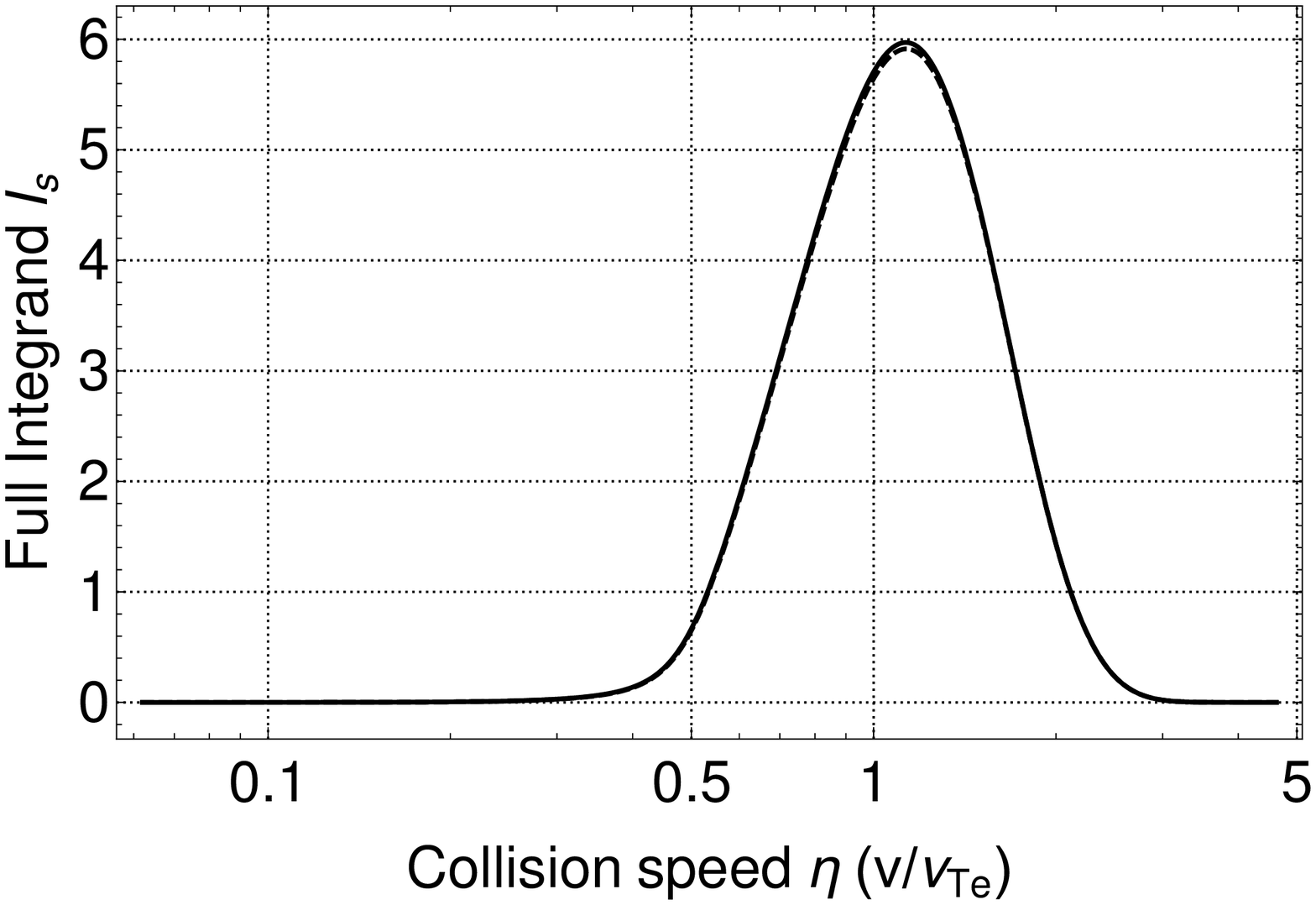}\\
\includegraphics[width=8.55cm]{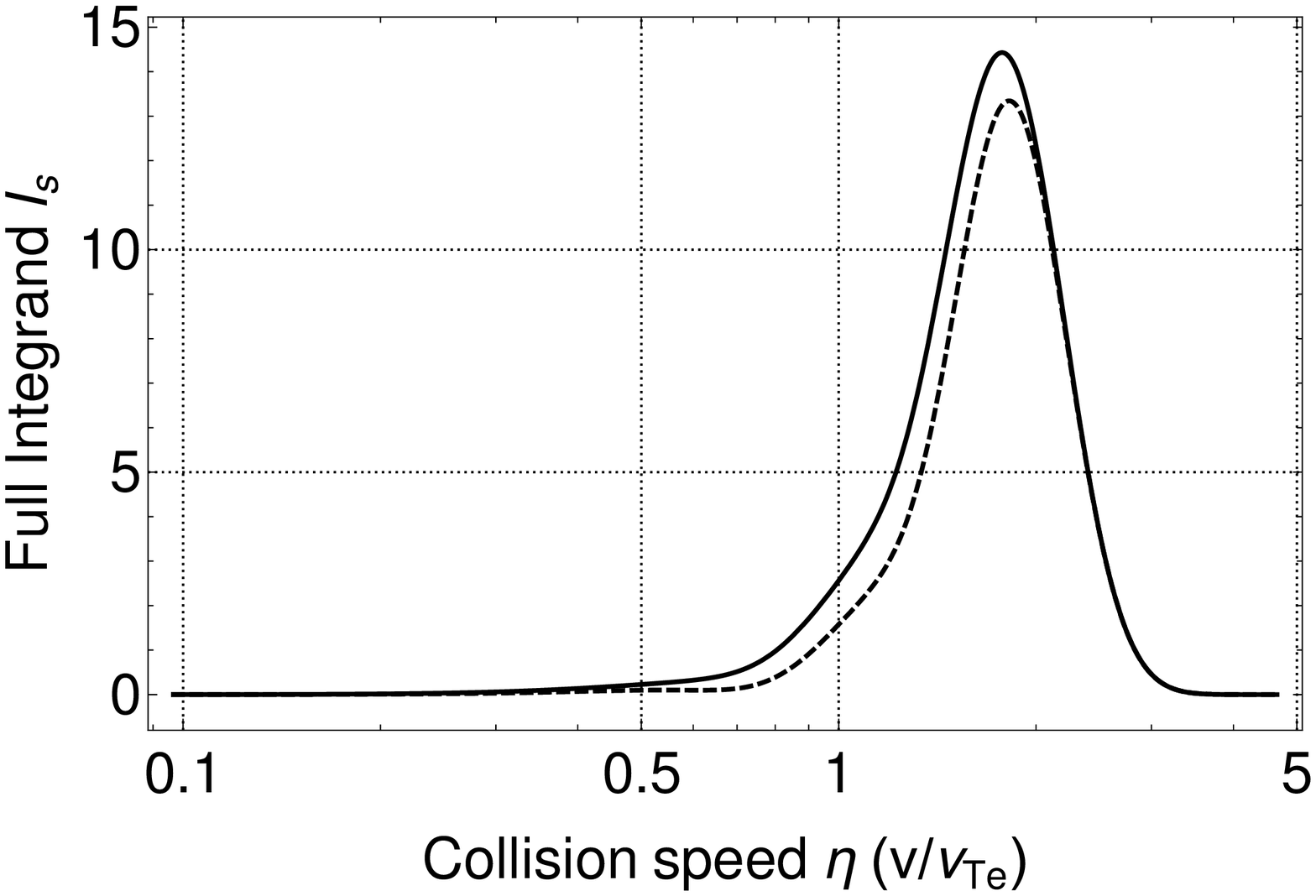}\\
\includegraphics[width=8.55cm]{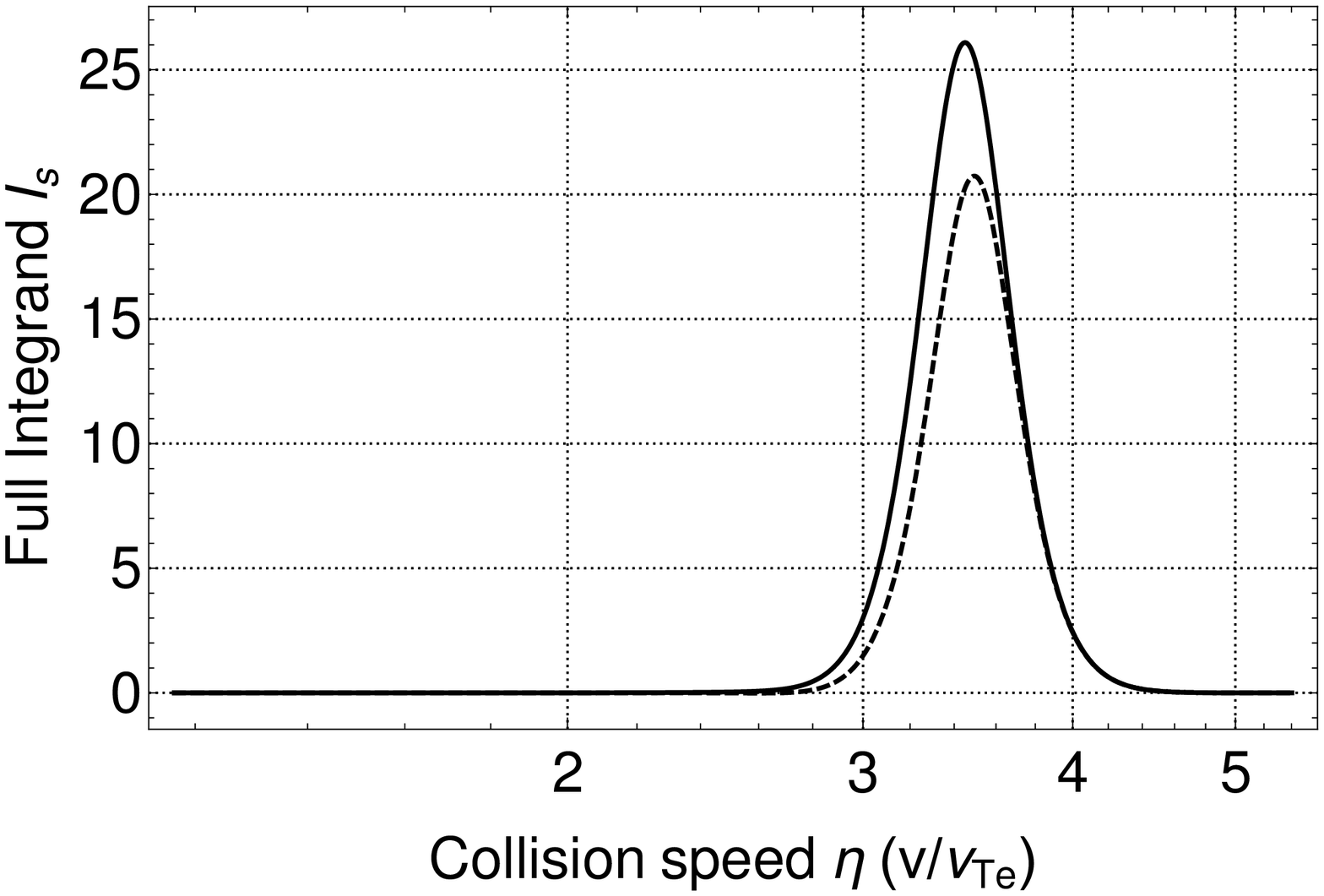}

\caption{\label{fig:momINT} Integrands ($I_{\epsilon}(\eta)$ solid and $I_{p}(\eta)$
dashed) appearing in equations (\ref{eq:xiE1}) and (\ref{eq:xiMOM1})
for aluminum at $2.7\thinspace{\rm g\cdot cm}^{-3}$ at three different
temperatures: $100\thinspace{\rm eV}$ ($\Theta\gg1$) (top), $10\thinspace{\rm eV}$
($\Theta\sim1$) (middle), $1\thinspace{\rm eV}$ ($\Theta\ll1$)
(bottom).}
\end{figure}

To illustrate the application of the model, we now turn to evaluating
it, with input potentials provided by the AA-TCP model (\citep{Starrett2012,Starrett2013})
for aluminum at a density of $2.7\thinspace{\rm g\cdot cm}^{-3}$,
over a range of temperatures spanning from the degenerate moderately
coupled to classical weakly-coupled regimes. However, firstly we demonstrate
the behaviors of the two functions $\sigma_{2}^{(1)}$ and $\sigma_{1}^{(1)}$
in figure \ref{fig:momCC} at two example temperature-density points.
The combined influence of the negative values of $\sigma_{2}^{(1)}$
and the preceding negative sign in equation (\ref{eq:xiMOM1}) leads
to interesting behavior in the integrand for the Coulomb integral.
The full integrand of equation (\ref{eq:xiMOM1}) is shown in figure
\ref{fig:momINT} where it is seen that the resulting integrals are
positive, as required. Note that the integrands are peaked functions;
broad and peaked near the thermal velocity $v_{Ts}$ in the classical
limit, and narrow with peak near the Fermi velocity in the degenerate
limit. Also note that $I_{p}$ and $I_{\epsilon}$ are identical in
the classical limit, but differ substantially in the degenerate case
due to the presence of the $\sigma_{2}^{(1)}$ factor.

\subsection{\label{subsec:deuterium}Relaxation Rates in Solid Density Aluminum
Plasma}

\begin{figure}[!tph]
\includegraphics[width=8.55cm]{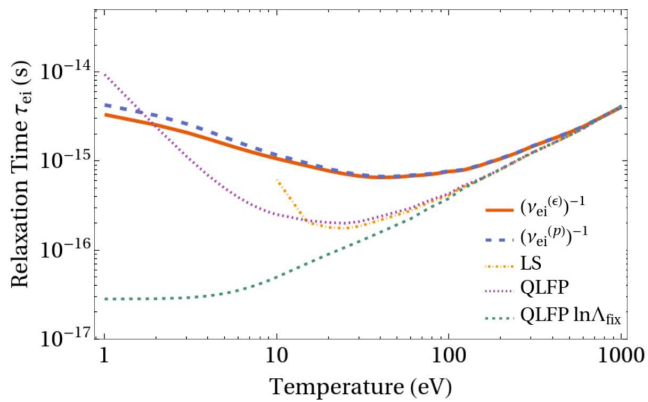}\caption{\label{fig:wide-1}Electron-ion collisional relaxation times ($\tau=\nu_{ei,}^{-1}$)
as a function of temperature in solid density ($2.7\thinspace{\rm g\cdot cm}^{-3}$)
aluminum.}
\end{figure}

Figure \ref{fig:wide-1} shows a comparison of the momentum and energy
relaxation rates. Each model is compared with the well-established
Landau-Spitzer result \citep{Spitzer1956}, which in the limit $m_{e}T_{i}\ll m_{i}T_{e}$
reduces to
\begin{equation}
\nu_{ei}^{{\rm LS}}\approx\nu_{0}\ln\Lambda_{{\rm LS}}\label{eq:nuLS}
\end{equation}
which has been verified in the classical limit given sufficiently
weak coupling \citep{Trintchouk2003,Kuritsyn2006}. The relaxation
rate predicted by the LFP model is(see equations 14-17 of \citep{Daligault2016a}),
\begin{equation}
\nu_{ei}^{{\rm LFP}}=\nu_{0}\left(\ln\Lambda_{{\rm LFP}}\frac{\xi}{1+\xi}\frac{3\sqrt{\pi}\Theta^{3/2}}{4}\right).\label{eq:nueilfp}
\end{equation}
We further note that our expression for the temperature relaxation
rate (given by equations (\ref{eq:Trelax})-(\ref{eq:momCC1})) is
the same as that recently obtained by a substantially different by
Daligault and Simoni (see equations (71)-(75) of \citep{Daligault2019})
if the potential of mean force is used for calculating the transport
cross section there. This equivalency can be seen through use of the
relation $n_{e}\left(h/\sqrt{\pi}m_{e}v_{Te}\right)^{3}=-2{\rm Li}_{3/2}\left(-\xi\right)$
from the normalization of the Fermi-Dirac distribution.

At a given density, as temperature decreases the Coulomb logarithm
will eventually reach zero due to neglect of strong coupling physics.
The resulting divergence of the Landau-Spitzer result due to the presence
of the (inverse) Coulomb logarithm
\begin{equation}
\ln\Lambda_{{\rm LS}}=\ln\frac{b_{{\rm max}}}{b_{{\rm min}}}.
\end{equation}
The maximum impact parameter is modeled as the larger of the screening
length $\lambda_{{\rm sc}}${[}equation (\ref{eq:rsc}){]} or the
Wigner-Seitz radius $a=(3/4\pi n_{i})^{1/3}$, and the minimum is
the larger of the classical distance of closest approach $r_{L}=e^{2}/k_{B}T$
or the thermal de Broglie wavelength $\lambda_{{\rm dB}}=\hbar/\left(m_{e}k_{{\rm B}}T_{e}\right)^{1/2}$
\citep{Lee1984}. In WDM, the vanishing Coulomb logarithm is often
resolved through the modification (see e.g. \citep{Lee1984})
\begin{equation}
\ln\Lambda_{{\rm LFP}}=\frac{1}{2}\ln\left(1+\frac{b_{{\rm max}}^{2}}{b_{{\rm min}}^{2}}\right),\label{eq:lambdaLFP}
\end{equation}
which we apply in our evaluation of the LFP model. This is often further
altered, as is done in the Lee-Moore conductivity model \citep{Lee1984},
by enforcing that the minimum value of the Coulomb logarithm be 2:
\begin{equation}
\ln\Lambda_{{\rm fix}}={\rm max}\left[2,\frac{1}{2}\ln\left(1+\frac{b_{{\rm max}}^{2}}{b_{{\rm min}}^{2}}\right)\right].\label{eq:lambdaLM}
\end{equation}
The approximations inherent in this approach are two-fold: small-angle
collisions must be assumed to obtain the LFP equation, and the choice
of maximum and minimum impact parameters represents an uncontrolled
expansion in the strongly coupled regime. The convergent kinetic equation
in our approach avoids these limitations.

Figure \ref{fig:wide-1} confirms the expectation that all expressions
agree at high temperatures associated with the the weakly-coupled
classical regime, while at low temperature the models differ as a
result of the different levels of inclusion of the physics of strong
coupling and degeneracy. In each case there is a minimum in the relaxation
time. In all cases except the Landau-Spitzer result, this minimum
can be attributed to a combination of both degeneracy and strong coupling:
strong coupling increases the collisionality of the system while the
onset of degeneracy reduces the collisionality through Pauli blocking.
The decreased level of ionization at lower temperatures also reduces
the collisionality. If the density is less than $10^{23}\thinspace{\rm cm}^{-3}$
as temperature is reduced the plasma will first become strongly coupled
and then degenerate, and if the density is greater than $10^{23}\thinspace{\rm cm}^{-3}$
the electrons will be degenerate when the transition to strong coupling
occurs.

The quantum mean force model predicts that the relaxation time $\tau$
for energy and momentum relaxation do not have the same behavior with
temperature. The rates are equal in the classical limit as expected,
but differ for lower temperatures when degeneracy arises. Generally,
the rates are smaller for energy relaxation, with the maximum difference
being a factor of $\sim2$. Experimental validation of this phenomenon
will require accurate measurements of both momentum and temperature
relaxation rates in WDM, a matter of considerable difficulty. However,
further consideration of the physical basis for the difference between
momentum and temperature relaxation is called for, and perhaps computational
methods will prove to be effective to this end. This effect is not
present in the LS or QLFP theories and is a result of allowing for
strong quantum collisions, which these theories do not account for.
The LS theory ignores both degeneracy and large-angle scattering.
The QLFP theory extends further into the degenerate regime and has
fixed the vanishing Coulomb log, but does not account for either correlations
or large-angle scattering when there is strong Coulomb coupling. The
divergence between the QLFP results using the two different prescriptions
for the Coulomb log illustrates the lack of strong-coupling physics
in the method.

\subsection{Electrical Conductivity of Solid-Density Aluminum Plasma}

We proceed to evaluate the electrical conductivity according to equation
(\ref{eq:conductivity}) for aluminum at $2.7\thinspace{\rm g\cdot cm}^{-3}$,
as a demonstration of the model in a regime marked by partial ionization
and a simultaneous transition from weak to strong coupling and classical
to degenerate statistics. For comparison we select the Lee-Moore model,
the model of Shaffer and Starrett \citep{Shaffer2020} and the QMD
simulations of Witte et al \citep{Witte2018}. The electrical conductivity
coefficient predicted by the LM model \citep{Lee1984} is
\begin{equation}
\sigma_{e}=\frac{ne^{2}}{m}\left\{ \frac{3\sqrt{m}(kT)^{3/2}}{2\sqrt{2}\pi Z^{2}n_{i}e^{4}\ln\Lambda_{{\rm fix}}}\right\} \frac{4}{3}\frac{\int_{0}^{\infty}\frac{t^{2}dt}{1+\exp(t-\mu/kT)}}{\int_{0}^{\infty}\frac{t^{1/2}dt}{1+\exp(t-\mu/kT)}}
\end{equation}
which we relate to the friction force density $\boldsymbol{R}$ and
thus the scattering rate: $\nu_{ei}^{(p)}=e^{2}n_{e}/\sigma m_{e}$
giving
\begin{equation}
\nu_{ei}^{{\rm LM}}=\nu_{0}\left[\ln\Lambda_{{\rm fix}}\frac{{\rm Li}_{3/2}(-\xi)}{{\rm Li}_{3}(-\xi)}\right].\label{eq:nulm}
\end{equation}
The Starrett and Shaffer model similarly uses the quantum potential
of mean force to mediate scattering, but in the context of the QLFP
equation. In order to introduce the effect of large-angle collisions
into the model they introduce a Coulomb logarithm defined via the
relaxation-time approximation (RTA) which we will refer to as ${\rm ln}\Lambda_{{\rm SS}}$.
For a commensurate comparison with our method (where we assume a Fermi
distribution for the electrons) and the QMD simulations, we neglect
the higher order Chapman-Enskog corrections associated with electron-electron
interactions that can be obtained in the SS model. The electron-ion
contribution corresponds with the first order of the Chapman-Enskog
expansion,
\begin{gather*}
\sigma_{{\rm 1,qLFP}}=\frac{3(4\pi\epsilon_{0})^{2}(k_{B}T)^{3/2}}{4\sqrt{2\pi m_{e}}Ze^{2}\ln\Lambda_{{\rm ss}}}\frac{{\rm Li_{3/2}(-\xi)}}{{\rm Li}_{0}(-\xi)}
\end{gather*}
With the identification of 
\[
{\rm Li}_{0}(-\xi)=\frac{-\xi}{1+\xi}
\]
and (from \ref{eq:nueilfp}) 
\[
\ln\Lambda\frac{\xi}{1+\xi}\frac{3\sqrt{\pi}\Theta^{3/2}}{4}\rightarrow\ln\Lambda,
\]
and equation (\ref{eq:xi_Theta_relation}) it can be seen this is
equivalent in form to equation (\ref{eq:conductivity}) with the difference
being the Coulomb logarithm.
\begin{figure}[!tph]
\includegraphics[width=8.55cm]{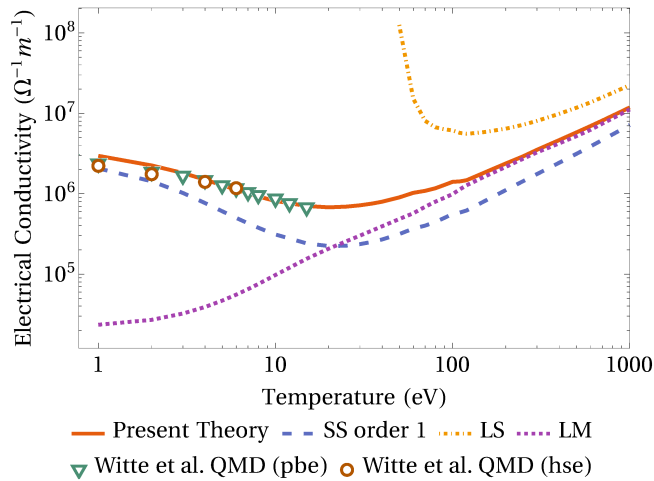}\caption{\label{fig:conductivity_aluminum}Electron-ion contribution to the
electrical conductivity of solid density aluminum ($2.7\thinspace{\rm g\cdot cm}^{-3}$)
as derived through the current work (solid line) the Starrett and
Shaffer model evaluated at first order in the Chapman-Enskog expansion
(dashed line), the Lee-More model (dotted line), and the LS conductivity
(dot-dashed line), along with QMD results of Witte et al \citep{Witte2018}
using the Perdew\textendash Burke-Ernzerhof and Heyd\textendash Scuseria\textendash Ernzerhof
exchange-correlation functionals.}
\end{figure}

The resulting predictions for the conductivity are shown in figure
\ref{fig:conductivity_aluminum}. Similarly to the relaxation times,
there is a minimum in the conductivity near the Fermi temperature.
This again can be attributed to both correlations and Pauli blocking
\citep{Shaffer2020}. Also as in the case of the relaxation times,
the LS theory fails to accurately predict the conductivity when degeneracy
and correlations are important, as expected. Furthermore, the commonly
used Lee-Moore theory performs poorly as a result of the correlations.
Interestingly, the Lee-Moore theory can be reproduced by replacing
the Coulomb logarithm in the standard QLFP formulation with the fixed
version prescribed in the Lee-Moore theory. Although we focus on the
electron-ion contribution, it is known that electron-electron interactions
cause a contribution of comparable magnitude in the classical weakly
coupled limit (the Spitzer correction) \citep{Spitzer1956}. However,
it can be expected that e-e collisions will be greatly suppressed
below the Fermi temperature due to Pauli blocking and therefore the
corrections due to a higher-order Chapman Enskog expansion will be
diminished at lower temperatures. Indeed, this is seen for the QLFP
equation \citep{Daligault2018}.

More interesting are the comparisons of the present theory with the
Shaffer-Starrett formulation of the QLFP theory \citep{Shaffer2020},
and with the QMD simulations of Witte et al. \citep{Witte2018}. The
QLFP equation being the limit of the BUU equation with only small-angle
scattering, it may be expected that these formulations should agree
in the limit of weak coupling. However, the present theory and the
SS theory differ in their treatment of the potential of mean force,
and the curves appear to not yet have reached this limiting behavior
at $1000\thinspace{\rm eV}$. The QMD simulations also make an interesting
direct comparison. QMD simulations do not directly include e-e collisions
because they use the Born-Oppenhiemer approximation, but account for
some level of the electronic interactions through the mean field.
\citep{Desjarlais2017}. Thus, it seems most appropriate to compare
the QMD simulations with theories evaluated to treat only the electron-ion
interactions, as is done in figure \ref{fig:conductivity_aluminum}.
Indeed, the agreement with these simulations is remarkable for most
of the range of available data, from $15\thinspace{\rm eV}$ down
to approximately $2\thinspace{\rm eV}$. At the lowest temperatures
the predictions begin to diverge, but it is unclear at these very
low temperatures whether the BUU equation can be expected to be valid
as higher-order quantum correlations come into play. To fully address
the contribution of e-e collisions will require solutions of the BUU
equation at higher orders of the Chapman-Enskog expansion.

The good agreement between QMD and the BUU predictions provides evidence
that large momentum transfer collisions, and the associated second
(quantum) contribution to the momentum scattering rate {[}equation
(\ref{eq:momCC2a}){]} are real and significant effects influencing
the electrical conductivity. This points to important physics beyond
what is captured by the QLFP theory, or its modifications, as is shown
by comparing with the first-order Chapman-Enskog solution of the Shaffer-Starrett
model from \citep{Shaffer2020} (the first order of this method is
equivalent to the electron-ion relaxation model described in the previous
section, and therefore provides a commensurate comparison). At the
same time, it is also important to note that electron-electron contributions
may influence the total conductivity at these conditions. Shaffer
and Starrett predict these to make order-unity contributions over
the range of conditions plotted in figure \ref{fig:conductivity_aluminum}
\citep{Shaffer2020}. Further development will be required to evaluate
this contribution from the BUU equation, as well as to provide a conclusive
test using QMD.

\section{\label{sec:conclusions}Conclusions}

We have presented a model for transport in plasmas with weak to moderate
Coulomb coupling and weak to moderate electron degeneracy. The model
is based on the quantum Boltzmann equation of Uehling and Uhlenbeck,
in which the two-body scattering is mediated by the equilibrium potential
of mean force. This incorporates correlations in the equilibrium limit
while maintaining the simplicity of binary collisions in the dynamical
equation. This is relevant to electron-ion collisions in WDM. As input
into the model, we utilized an existing model for the potential of
mean force derived from the quantum Ornstein-Zernike equations and
an average-atom quantum hypernetted-chain-approximation model \citep{Starrett2012,Starrett2013,Starrett2018}.

The model was used to compute momentum and energy relaxation rates.
The transport coefficients were written analogously to the classical
Landau-Spitzer (LS) result in terms of a ``Coulomb integral'' that
takes the place of the traditional Coulomb logarithm. The Coulomb
integral depends on the level of degeneracy, and Coulomb coupling
enters through the calculation of the momentum-transfer cross section
solving the Schr\"{o}dinger equation with the PMF as the scattering
potential. The momentum relaxation rate was found to differ from temperature
relaxation in that it depends on a different transport cross section,
which includes a term that is solely associated with degeneracy, and
has no analog in the classical limit. The dependence of the integrands
of the Coulomb integrals on the level of degeneracy was compared for
the temperature and momentum relaxation cases.

We concluded by calculating the temperature and momentum relaxation
rates and electrical conductivity in solid density aluminum plasma
over a range of temperatures that covered the transitions between
weak and moderate coupling and weak and moderate degeneracy. Predictions
were compared with other leading models. It was found that all models
behave as expected in the classical weak-coupling limit, and diverge
widely in the limit of a degenerate moderately-coupled plasma. We
assessed the relative importance of the different relevant physical
processes that complicate the problem as degeneracy and coupling simultaneously
increase: diffraction, Pauli blocking, correlations, and large-angle
scattering. Interestingly, in the degenerate regime there is a quantitative
difference in the predicted relaxation rates for momentum versus energy.
Ultimately, current and near-future experimental measurements \citep{Cho2016,Glenzer2016,Zaghoo2019}
and ab-initio simulations \citep{Daligault2018a,Larder2019,Ma2019,Simoni2019,White2020,Yilmaz2020}
will be able to shed light on the applicability of the different models
of transport for WDM.

This work can be improved through inclusion of electron-electron collisions
and higher-order terms of a Chapman-Enskog expansion. Additionally,
further work will be required to obtain a rigorously derived convergent
kinetic equation with the appropriate potential of mean force. Finally,
recent and upcoming experimental measurements of electrical conductivity
and temperature relaxation \citep{Cho2016,Zaghoo2019} may soon open
the door for discrimination between the validity of the various models
of relaxation in WDM. This will enhance our understanding of the basic
physics of WDM, and allow increased fidelity in the rapid calculation
of transport coefficients for use in hydrodynamic simulations of naturally
and experimentally occurring WDM.
\begin{acknowledgments}
The authors wish to acknowledge Charles Starrett and Nathaniel Shaffer
for the provision of input data at equilibrium for the potential of
mean force and for their valuable comments on this work. This material
is based upon work supported by the U.S. Department of Energy, Office
of Science, Office of Fusion Energy Sciences under Award Number DE-SC0016159.
\end{acknowledgments}

\appendix

\section{\label{sec:phaseshifts}Determination of Phase Shifts}

Solution of the scattering problem comes down to solution of the radial
Schr\"{o}dinger equation \citep{Landau1965}
\[
\frac{d^{2}u_{l}}{dr^{2}}+\left[k^{2}-\frac{l\left(l+1\right)}{r^{2}}-\frac{2m_{ei}}{\hbar^{2}}W^{(2)}\left(r\right)\right]u_{l}=0,
\]
For each angular quantum number $l$ there is a phase shift $\delta_{l}$
that can be extracted from the asymptotic behavior of the wavefunction
$u_{l}$ beyond the range of the potential at point $R$ (defined
as a point beyond with the influence of the potential on the wavefunction
is negligible) through the relation:
\[
\tan\delta_{l}=\frac{kRj_{l}'\left(kR\right)-\beta_{l}j_{l}\left(kR\right)}{kRy_{l}'\left(kR\right)-\beta_{l}y_{l}\left(kR\right)}
\]
with
\[
\beta_{l}=\frac{1}{u_{l}/r}\left.\frac{d\left(u_{l}/r\right)}{dr}\right|_{r=R}
\]
where $j_{l}$ ($y_{l}$) are the spherical Bessel (Neumann) functions.
For $l>30$ it is faster and still accurate to use the WKB phase shifts
\begin{gather}
\delta_{l}^{({\rm WKB)}}=-\int_{(l+1/2)/k}^{\infty}\sqrt{k^{2}-\frac{(l+1/2)^{2}}{r^{2}}dr}\nonumber \\
+\int_{r_{C}}^{\infty}\sqrt{k^{2}-\frac{(l+1/2)^{2}}{r^{2}}-\frac{2m_{e}}{\hbar^{2}}U(r)}dr
\end{gather}

\section{\label{sec:shifts_appendix}Simplification of Cross Sections}

Cross sections are calculated in the partial wave expansion
\begin{equation}
\frac{d\sigma}{d\Omega}=\left|\frac{1}{2ik}\sum_{l=0}^{\infty}\left(2l+1\right)\left({\rm e}^{2i\delta_{l}}-1\right)P_{l}\left({\rm cos}\theta\right)\right|^{2},
\end{equation}
where the phase shifts $\delta_{l}$ are calculated from solution
of the Schr\"{o}dinger equation for the given potential. This can
be written as a double sum
\begin{gather*}
\frac{d\sigma}{d\Omega}=\frac{1}{k^{2}}\sum_{n=0}^{\infty}\sum_{m=0}^{\infty}\left(2m+1\right)\left(2n+1\right)\\
\times{\rm e}^{i\left(\delta_{m}-\delta_{n}\right)}\sin\delta_{m}\sin\delta_{n}{\rm P}_{m}(\cos\theta){\rm P}_{n}(\cos\theta)
\end{gather*}
which inserted into equation (\ref{eq:momCC2a}) results in
\begin{gather*}
\sigma_{2}^{(1)}\left(\eta,\Gamma\right)=\frac{4\pi}{k^{2}}\\
\times\sum_{n=0}^{\infty}\sum_{m=0}^{\infty}\left(2m+1\right)\left(2n+1\right){\rm e}^{i\left(\delta_{m}-\delta_{n}\right)}\sin\delta_{m}\sin\delta_{n}\\
\times\int_{0}^{\pi}d\theta\sin^{2}\frac{\theta}{2}\sin\theta\cos\theta{\rm P}_{m}(\cos\theta){\rm P}_{n}(\cos\theta).
\end{gather*}
Taking advantage of the properties of the Legendre polynomials this
is
\begin{gather*}
\frac{4\pi}{k^{2}}\sum_{n=0}^{\infty}\sum_{m=0}^{\infty}\left(2m+1\right)\left(2n+1\right){\rm e}^{i\left(\delta_{m}-\delta_{n}\right)}\sin\delta_{m}\sin\delta_{n}\\
\times\frac{1}{2}\int_{0}^{\pi}\sin\theta\left[P_{1}(\cos\theta)-\frac{2}{3}P_{2}(\cos\theta)-\frac{P_{0}}{3}\right]\\
\times P_{m}(\cos\theta)P_{n}(\cos\theta)d\theta,
\end{gather*}
Finally, with the identity
\begin{equation}
\int_{0}^{\pi}d\theta\sin\theta P_{l}(\cos\theta)P_{m}(\cos\theta)P_{n}(\cos\theta)=2\left(\begin{array}{ccc}
l & m & n\\
0 & 0 & 0
\end{array}\right)^{2},
\end{equation}
where $\left(\begin{array}{c}
\cdots\\
\cdots
\end{array}\right)$ is the Wigner 3j symbol and using the specific values for $l=0,\thinspace1,\thinspace{\rm and}\thinspace2$,
we obtain equation (\ref{eq:momCC2b}).

\bibliographystyle{apsrev4-1}
\bibliography{refs}

\end{document}